\documentclass[a4paper,amsmath,amssymb,twocolumn,superscriptaddress]{revtex4}

\usepackage{doi}
\usepackage{hyperref}
\hypersetup{
  colorlinks=true,        
  linkcolor=blue,         
  citecolor=cyan,         
}

\usepackage{graphicx}
\usepackage{dcolumn}
\usepackage{bm}
\usepackage{color}
\usepackage{enumitem}

\newcommand{\beq}{\begin{equation}}
\newcommand{\eeq}{\end{equation}}
\newcommand{\bea}{\begin{eqnarray}}
\newcommand{\eea}{\end{eqnarray}}

\usepackage{bbm}
\usepackage{amsfonts}
\usepackage{mathrsfs}
\usepackage{latexsym}
\usepackage{epsfig}
\usepackage{epstopdf}
\usepackage{dcolumn}
\usepackage{bm}
\usepackage{color}
\usepackage{comment}
\usepackage{xcolor}
\usepackage{orcidlink}
 \usepackage{float, geometry,lipsum}
\begin{document}

\title{Probing Quantum Gravity through Chaotic Orbits and Strong-Field Effects in Kerr Black Holes Embedded in Perfect Fluid Dark Matter}


\author{Shubham Kala}
\email{shubhamkala871@gmail.com}
\affiliation{The Institute of Mathematical Sciences, C.I.T. Campus,
Taramani, Chennai--600113, Tamil Nadu, India}
\affiliation{Research Center of Astrophysics and Cosmology, Khazar University, Baku, AZ1096, 41 Mehseti Street, Azerbaijan}

\author{Sara Saghafi}
\email{saghafisara1366@gmail.com}
\affiliation{Department of Theoretical Physics, Faculty of Science, University of Mazandaran, P. O. Box 47416-95447, Babolsar, Iran}

\author{M. Yousaf}
\email{myousaf.math@gmail.com}
\affiliation{Department of Mathematics, Virtual University of Pakistan,\\ 54-Lawrence Road, Lahore 54000, Pakistan.}

\author{Hemwati Nandan}
\email{hnandan@associates.iucaa.in}
\affiliation{Department of Physics, Hemwati Nandan Bahuguna Garhwal University,
Srinagar--246174, Uttarakhand, India}

\author{Ahmadjon~Abdujabbarov}
\email{ahmadjon@astrin.uz}
\affiliation{School of Physics, Harbin Institute of Technology, Harbin 150001, People’s Republic of China}

\author{Chengxun Yuan}
   \email{yuancx@hit.edu.cn (Corresponding Author)}
\affiliation{School of Physics, Harbin Institute of Technology, Harbin 150001, People’s Republic of China}

\author{G. Mustafa}
\email{gmustafa3828@gmail.com (Corresponding Author)}
\affiliation{Department of Physics, Zhejiang Normal University, Jinhua 321004, People's Republic of China}

\begin{abstract}
We study the nonlinear photon dynamics in quantum improved rotating black hole surrounded by perfect fluid dark matter (PFDM) using several methods of analysis, including Poincaré sections, Lyapunov exponents, Kolmogorov-Sinai (KS) entropy and weighted Birkhoff averages (WBA). In particular, we examine the effect of quantum-improved parameter $(\tilde{\omega})$ and PFDM parameter $(\zeta)$ on the null geodesic motion hence stability of circular orbit. Poincaré sections illustrate the transition from regular to chaotic motion as these parameters increase, characterized by the deformation and fragmentation of invariant tori and the emergence of scattered chaotic regions in phase space. The stability properties of null circular orbits are quantified through the Lyapunov indicators, revealing that both the quantum improvement parameter and the PFDM parameter enhance the sensitivity of photon trajectories to initial conditions. The KS entropy provides an independent measure of dynamical complexity and confirms the growth of chaotic behavior with increasing quantum and PFDM corrections. Additionally, the WBA method offers a robust quantitative criterion for distinguishing regular and chaotic orbits and allows a detailed mapping of the phase-space structure. The results demonstrate that the combined effects of quantum gravity corrections and PFDM significantly modify the effective potential governing photon motion, leading to a rich mixed phase-space structure with coexisting regular and chaotic regions. These findings underscore the crucial role of quantum and dark matter contributions in shaping photon dynamics near rotating black holes and suggest possible observational implications on black hole shadows and gravitational lensing in strong-field regimes.
\end{abstract}

\maketitle

\date{\today}


\section{Introduction}\label{intro}

Black hole thermodynamics represents a fascinating interface between general theory of relativity (GTR),
quantum theory, statistical physics, and information theory, and it offers deep clues about
the fundamental nature of gravity. One of the earliest indications of this connection comes
from the area theorem, which states that the total event horizon area of a black hole never
decreases with time \cite{hawking1971gravitational} which closely resembles the
second law of thermodynamics and motivated the interpretation that black holes possess
thermodynamic attributes. Based on this analogy, Bekenstein proposed that the entropy
of a black hole is proportional to the area of its event horizon \cite{bekenstein1973black,tsallis2013black}.
The thermodynamic interpretation strengthened when Hawking showed that black
holes emit thermal radiation, implying that they have a definite temperature
\cite{hawking1974black}, while black hole explosions as probes of new physics studied in \cite{federico2025black}. 
This discovery established a consistent correspondence
between the laws of black hole mechanics and ordinary thermodynamics.

Later, Hawking and his collaborator demonstrated that Schwarzschild black holes in an anti-de
Sitter background undergo a phase transition between a black hole configuration
and thermal radiation \cite{hawking1975particle}, while thermodynamics of black holes 
in anti-de Sitter space investigated in \cite{hawking1983thermodynamics}. The development of the
anti-de Sitter/CFT correspondence further motivated detailed studies of the thermodynamic
behavior and phase structure of anti-de
Sitter black holes \cite{maldacena1999large}.
In particular, Reissner-Nordström anti-de
Sitter black holes exhibit behavior closely
analogous to a van der Waals fluid, whereas in the canonical ensemble they show a first-order
phase transition that terminates at a second-order critical point,
while in the grand canonical ensemble a Hawking Page type transition appeared in \cite{cvetic1999phases}. 
Caldarelli and his collaborators \cite{caldarelli2000thermodynamics} investigated the thermodynamic properties of 4D Kerr-Newman-anti-de Sitter black holes in both canonical and grand-canonical ensembles, while the stability analysis
reveals a Hawking-type transition in the grand canonical ensemble. In the
canonical ensemble a first-order phase transition occurs between small and large black
holes, ending at a critical point for sufficiently large charge or angular momentum.

In GTR, the gravitational field surrounding a black hole is interpreted
as the curvature of spacetime, while this theory predicted the formation of singularities as well as
determined the position of the event horizon through parameters such as the black hole
spin. Nevertheless, GTR also leads to regions where the classical description fails,
particularly at the spacetime singularity where known physical laws cease to be valid, whereas this limitation indicates the necessity of incorporating quantum effects into the
gravitational regime. Key phenomena associated with black holes including spacetime singularities, event
horizons, and Hawking radiation \cite{hawking1975particle} cannot be completely
understood within purely classical or semiclassical treatments. Consequently, addressing
the singularity problem, explaining black hole evaporation and the information loss
paradox, as well as establishing a consistent formulation of black hole thermodynamics all
point toward the requirement of a full theory of quantum gravity. Among the different candidates for a quantum theory of gravity, loop quantum gravity provides a non-perturbative and background independent quantization of spacetime geometry, where geometric observables acquire discrete spectra, however, When this framework is implemented in spherically symmetric configurations, the resulting effective metrics describe loop
quantum black holes. These models remove the classical central singularity and
replace it with a regular, non-singular core structure (see for details  \cite{modesto2004disappearance,modesto2010semiclassical,bambi2013non,ashtekar2018quantum}).

Regular (non-singular) black hole models have been extensively studied in the literature, with comprehensive reviews provided in Refs.~\cite{torres2022regular,lan2023regular}. Among these, the 
formation and evaporation of regular black holes studied by Hayward in Ref.~\cite{hayward2012formation} represents a notable example. Beyond regular black holes, a wide range of alternative black hole configurations have been explored in the literature. These include Einstein-aether solutions, black holes surrounded by quintessence, the Kerr-Newman-Kasuya spacetime, and Kerr solutions in modified gravity, along with several other related 
models~\cite{moffat2015black,Kala:2020prt,Kala:2021ppi,wang2022optical,gao2023investigating,yousaf2024fuzzy,Roy:2025qmx}. Chen and his collaborators \cite{chen2024quantum} examined quantum corrected Kerr black holes within  asymptotically safe gravity by employing mass dependent scale identifications, and found that a physically consistent identification must depend on the combination $Mr$ and on the horizon area. Using a representative choice of this identification, they analyzed the properties of the resulting rotating quantum black holes, while modified regular solutions and possessed several appealing features, like a well-defined thermodynamic description at the horizon, removal of the classical Kerr ring singularity, and a partial suppression of the closed timelike curves that occur in the classical Kerr geometry. 

Motivated by the expectation that both quantum effects and astrophysical environments can modify the classical Kerr picture, we consider a rotating black hole geometry that incorporates first quantum-gravity inspired corrections and secondly the influence of perfect fluid dark matter (PFDM), while PFDM sector is modeled through an anisotropic effective fluid description that is frequently employed to represent dark matter surrounding compact entities \cite{jana2026quasi}. In this scenario, the PFDM parameter regulates the strength of the environmental contribution and induces a characteristic
logarithmic modification in the gravitational potential, thereby altering the causal and optical
properties of the spacetime compared to the vacuum Kerr solution.
To account for quantum effects in a tractable way, we adopt an RG improved prescription in which
the gravitational coupling effectively runs with scale, whereas such approach encapsulates quantum
corrections expected in strong-field regions and can be encoded through an effective mass profile
that simultaneously captures the PFDM contribution and the quantum improvement. Consequently, the near-core behavior is softened relative to the classical Kerr spacetime, and the combined
effects of rotation, PFDM, and quantum corrections can yield a richer horizon phenomenology.

The study of black hole spacetimes modified by quantum gravitational 
corrections and dark matter distributions has attracted considerable attention in recent years. In this context, Yildiz et al.~\cite{Yildiz:2024dkt} investigated the optical properties of the 
Euler--Heisenberg black hole surrounded by perfect fluid dark matter, demonstrating that the PFDM distribution produces systematic modifications to the photon sphere radius, shadow angular diameter, and gravitational lensing observables, providing a direct benchmark for the present analysis. The astrophysical implications of dark matter environments were further examined by Ashraf et al.~\cite{Ashraf:2025nvt}, who derived observational constraints on 
quasi-periodic oscillations around a charged non-commutative Schwarzschild black hole surrounded by PFDM, establishing that the dark matter parameter induces measurable shifts in the epicyclic 
frequency ratios resolvable by current X-ray timing facilities, a result directly relevant to the chaotic breakdown of orbital resonances identified in the present work. Caliskan et 
al.~\cite{Caliskan:2024rdu} further analyzed particle trajectories and epicyclic oscillations around a piece-wise black hole immersed in dark matter, providing a systematic characterization of how matter distributions modify the effective potential structure and ISCO location in the strong-field regime. The present manuscript extends these investigations to the nonlinear phase-space structure of null geodesics in a quantum-improved Kerr black hole surrounded by perfect 
fluid dark matter, connecting chaotic photon dynamics to broader observational signatures in modified black hole spacetimes. Recent studies have revealed the emergence of nonlinear phenomena, such as fractal structures, multifractal chaos, and irregular photon trajectories, in various relativistic settings \cite{de2021fractal,oliveira2025multifractal}.

In order to obtain a rotating extension of the PFDM corrected seed geometry, we employ a modified Newman-Janis type construction that enforces a manifestly real rotating metric and reproduces the correct Kerr limit~\cite{azreg2014generating,azreg2014static}. Upon transforming to Boyer-Lindquist type coordinates, the resulting axisymmetric spacetime is characterized by the usual rotation parameter together with two extra parameters controlling the PFDM 
background and the quantum improvement. The horizon structure then follows from the Kerr like horizon function, which generally becomes transcendental once the PFDM induced logarithmic behavior is present. Having established the background geometry, we investigate its dynamical implications through null geodesic motion and the stability of photon orbits, since the spacetime is stationary and axisymmetric, photon trajectories admit conserved quantities, enabling a systematic analysis of circular photon orbits and their instability using Lyapunov type diagnostics~\cite{Wu:2006rx,Han:2008zzf,Chen:2016tmr}. 

Phase-space techniques such as Poincar\'{e} sections are widely used to characterize the transition from regular to chaotic behavior and to distinguish ordered and chaotic dynamics in relativistic systems~\cite{Cornish:1996de,Guo:2022kio,Kumara:2024obd,Gallo:2024wju,Singh:2026vfd,Singh:2026odr}. In the present work, we employ these techniques alongside the Lyapunov exponent to provide a comprehensive characterization of the phase-space structure of null geodesic motion in the 
quantum-improved Kerr black hole surrounded by PFDM. We further quantify chaos using fast indicators and entropy based measures; in particular, the Kolmogorov-Sinai entropy is closely connected to positive Lyapunov exponents via Pesin type relations
\cite{pesin1977characteristic,Pradhan:2015aaa,Mondal:2021exj,Singh:2026rbz}. Finally, for efficient global scans of phase space, weighted Birkhoff-average methods provide robust discrimination between regular and chaotic trajectories \cite{eckmann1985ergodic,cornfeld2012ergodic,das2016measuring,
meiss2021birkhoff,sander2020birkhoff,duignan2023distinguishing,rolim2025pynamicalsys,Kala:2026qej}. Chaos and stability properties in gravitational and dynamical systems are commonly characterized through tools such as Lyapunov exponents and entropy based measures, which quantify the sensitivity of trajectories to initial conditions \cite{wu2003computation,cornish2003lyapunov,deich2024lyapunov}. The coordinate invariant nature of relativistic chaos emphasized in the literature \cite{motter2003relativistic}, moreover, chaotic dynamics appear in broader contexts including holographic models and nonlinear Hamiltonian planetary systems \cite{basu2012chaos,el2026stability}, further highlight the robustness of chaotic behavior \cite{muni2026robust}.

The structure of our manuscript is as follows: In Section~\ref{S2}, we derive a quantum-impoved rotating black hole solution surrounded by PFDM. In Section~\ref{S3}, we derive the equations of motion for null geodesics. Section~\ref{S4} is devoted to the analysis of null circular orbits and their stability using Lyapunov exponents. In Section~\ref{S5}, we investigate the phase-space dynamics through Poincaré sections. Section~\ref{S6} introduces Lyapunov-based chaos indicators, including the LLE and the FLI. In Section~\ref{S7}, we analyze the chaotic behavior using the Kolmogorov–Sinai entropy. Section~\ref{S8} is dedicated to discrete chaos indicators based on the Weighted Birkhoff Average (WBA) method. Finally, Section~\ref{S9} summarizes our main results and provides concluding remarks and discussions. 

\section{Quantum Improved Kerr Black hole surrounded by PFDM}\label{S2}
The derivation of the metric begins with the action in the presence of PFDM~\cite{kiselev2003quintessence,kiselev2003quintessential},
\begin{equation}\label{eq1}
I = \int d^4x \, \sqrt{-g} \left( \frac{R}{16\pi G} + L_{\rm PFDM} \right),
\end{equation}
where \(R\) is the Ricci scalar, \(G\) is Newton's constant, and \(L_{\rm PFDM}\) is the Lagrangian describing the PFDM. The energy-momentum tensor for PFDM is taken as \cite{das2022study,xu2018kerr,das2024stability}
\begin{equation}\label{eq2}
(T^\mu_\nu)_{\rm PFDM} = \text{diag}(-\rho, P_r, P, P), \quad P_r = -\rho,
\end{equation}
with the equation of state \(P/\rho = \epsilon - 1 = 1/2\). Adopting a spherically symmetric ansatz for the metric \cite{das2022study},
\begin{equation}\label{eq3}
ds^2 = -e^{\sigma(r)} dt^2 + e^{\gamma(r)} dr^2 + r^2 (d\theta^2 + \sin^2\theta d\phi^2),
\end{equation}
the field equations yield \(\sigma + \gamma = 0\). Defining
\begin{equation}\label{eq4}
\sigma(r) = \ln(1 - B(r)) \quad \Rightarrow \quad f(r) = e^{\sigma(r)} = 1 - B(r),
\end{equation}
the solution of the differential equation for \(B(r)\) gives
\begin{equation}\label{eq4}
B(r) = \frac{r_s}{r} - \frac{\zeta}{r} \ln \frac{r}{ |\zeta|},
\end{equation}
where \(r_s = 2GM\) is the Schwarzschild radius and \(\zeta\) is the PFDM parameter. Applying the Renormalization Group (RG) improvement via a running Newton constant \cite{Bonanno:2000ep,harst2011qed,eichhorn2018upper,ishibashi2021quantum},
\begin{equation}\label{eq5}
G(r) = \frac{G}{1 + \tilde{\omega} G/r^2},
\end{equation}
the quantum-corrected lapse function becomes \cite{jana2026quasi}
\begin{equation}\label{eq6}
f(r) = 1 - \frac{2 G(r) M}{r} + \frac{\zeta}{r} \ln \frac{r}{ |\zeta|} = 1 - \frac{2GM}{r + \tilde{\omega} G} + \frac{\zeta}{r} \ln \frac{r}{ |\zeta|}.
\end{equation}
It is convenient to rewrite the seed metric in terms of an effective mass function $m(r)$ defined by
\begin{equation}
f(r)=1-\frac{2m(r)}{r},
\end{equation}
which gives
\begin{equation}
\frac{2m(r)}{r}=\frac{2GMr}{r^2+\tilde{\omega}G}-\frac{\zeta}{r}\ln\!\left(\frac{r}{|\zeta|}\right),
\end{equation}
hence
\begin{equation}
2m(r)=\frac{2GMr^2}{r^2+\tilde{\omega}G}-\zeta\ln\!\left(\frac{r}{|\zeta|}\right),\;
m(r)=\frac{GMr^2}{r^2+\tilde{\omega}G}-\frac{\zeta}{2}\ln\!\left(\frac{r}{|\zeta|}\right).
\end{equation}
To implement the advanced (modified) Newman--Janis algorithm \cite{azreg2014generating,azreg2014static} in a form free of ambiguities associated with complexification, one first passes to advanced Eddington--Finkelstein coordinates $(u,r,\theta,\phi)$ by introducing the null coordinate
\begin{equation}
du=dt-\frac{dr}{f(r)}\qquad \left(u=t-\int \frac{dr}{f(r)}\right),
\end{equation}
so that the seed metric becomes
\begin{equation}
ds^2=-f(r)\,du^2-2\,du\,dr+r^2d\theta^2+r^2\sin^2\theta\,d\phi^2.
\end{equation}
A standard Newman--Penrose null tetrad compatible with this geometry is chosen as
\begin{equation}
l^\mu=\delta^\mu_r,\qquad
n^\mu=\delta^\mu_u-\frac{f(r)}{2}\,\delta^\mu_r,
\end{equation}
\begin{equation}
m^\mu=\frac{1}{\sqrt2\,r}\left(\delta^\mu_\theta+\frac{i}{\sin\theta}\,\delta^\mu_\phi\right),\qquad
\bar m^\mu=\frac{1}{\sqrt2\,r}\left(\delta^\mu_\theta-\frac{i}{\sin\theta}\,\delta^\mu_\phi\right),
\end{equation}
which satisfies $g^{\mu\nu}=-l^{(\mu}n^{\nu)}+m^{(\mu}\bar m^{\nu)}$. The Newman--Janis complex transformation is then performed via
\begin{equation}
u\rightarrow u'=u-ia\cos\theta,\quad
r\rightarrow r'=r+ia\cos\theta,\quad
\theta'=\theta,\quad \phi'=\phi,
\end{equation}
and the prime notation is subsequently omitted for brevity. Within the framework of the modified Azreg-A\"{\i}nou–type prescription \cite{azreg2014generating,azreg2014static}, rather than postulating an ad hoc complexified expression for $f(r)$, one ensures both the reality of the metric functions and the correct Kerr limit by promoting the radial function to a real function of two variables and by consistently modifying the angular sector according to
\begin{equation}
\Sigma(r,\theta)=r^2+a^2\cos^2\theta,\qquad
F(r,\theta)=1-\frac{2m(r)\,r}{\Sigma}.
\end{equation}
With these identifications, the rotating geometry in advanced Eddington--Finkelstein--like coordinates takes the Kerr-type form
\begin{equation}
\begin{aligned}
ds^2=&-F\,du^2-2\,du\,dr+\Sigma\,d\theta^2
+\sin^2\theta\,(r^2+a^2)\,d\phi^2
\\&+2a\sin^2\theta\,dr\,d\phi
-2a\sin^2\theta\,(1-F)\,du\,d\phi,
\end{aligned}
\end{equation}
where $a$ is the rotation parameter. To obtain Boyer--Lindquist coordinates $(t,r,\theta,\varphi)$, one introduces the radial redefinitions
\begin{equation}
du=dt-\lambda(r)\,dr,\qquad d\phi=d\varphi-\chi(r)\,dr,
\end{equation}
and chooses $\lambda(r)$ and $\chi(r)$ so that the mixed terms $dt\,dr$ and $d\varphi\,dr$ vanish. Defining
\begin{equation}
\Delta(r)=r^2+a^2-2m(r)\,r,
\end{equation}
the standard Kerr-like choice
\begin{equation}
\lambda(r)=\frac{r^2+a^2}{\Delta(r)},\qquad
\chi(r)=\frac{a}{\Delta(r)}
\end{equation}
indeed cancels the unwanted cross terms and yields the final axisymmetric rotating metric in Boyer--Lindquist form,
\begin{equation}
\begin{aligned}
ds^2=&-\left(1-\frac{2m(r)\,r}{\Sigma}\right)dt^2
-\frac{4a\,m(r)\,r\sin^2\theta}{\Sigma}\,dt\,d\varphi
+\frac{\Sigma}{\Delta}\,dr^2\\
&\quad+\Sigma\,d\theta^2 +\sin^2\theta\left(r^2+a^2+\frac{2a^2m(r)\,r\sin^2\theta}{\Sigma}\right)d\varphi^2,
\end{aligned}
\end{equation}
with
\begin{equation}
\begin{aligned}
&\Sigma=r^2+a^2\cos^2\theta,\qquad
\Delta=r^2+a^2-2m(r)\,r,\qquad
\\&m(r)=\frac{GMr^2}{r^2+\tilde{\omega}G}-\frac{\zeta}{2}\ln\!\left(\frac{r}{|\zeta|}\right).
\end{aligned}
\end{equation}
In this rotating generalization, \(M\) denotes the mass parameter (reducing to the Kerr solution in the limit \(\tilde{\omega}=\zeta=0\)), \(a\) represents the specific angular momentum, \(\tilde{\omega}\) governs the renormalization-group (RG) improvement via the modification factor \(r^{2}+\tilde{\omega}G\), and \(\zeta\) characterizes the contribution of PFDM through the logarithmic term. The dimensionless parameter $\tilde{\omega}$ encodes the leading quantum gravitational correction arising from the non-perturbative fixed point structure of the gravitational renormalization group flow~\cite{Bonanno:2000ep}. In the limit $\tilde{\omega} \to 0$, the standard Kerr geometry is recovered exactly, whereas nonzero $\tilde{\omega}$ smooths out the classical central singularity and modifies the near-horizon geometry in a manner consistent with asymptotic safety scenarios of quantum gravity. The parameter $\tilde{\omega}$ is theoretically constrained to be non-negative, and while a precise observational bound has not yet been established from first principles, estimates based on the asymptotic safety program place it in the range $\tilde{\omega} \in [0, 1]$ in Planck units. Constraints from black hole shadow observations and gravitational lensing have further restricted $\tilde{\omega}$ to sub-unity values consistent with current Event Horizon Telescope measurements~\cite{Kala:2026bmz}. The PFDM parameter $\zeta$ characterizes the strength of the dark matter distribution surrounding the black hole and carries units of length in natural units. For $\zeta > 0$, the dark matter contribution adds a positive mass component at intermediate radii, effectively deepening the gravitational potential well and extending the influence of the black hole to larger distances, consistent with the flat rotation curves observed in galactic dynamics. For $\zeta < 0$, the correction acts in the opposite sense, reducing the effective mass at large $r$. The physically motivated range of $\zeta$ is typically taken as $|\zeta| \lesssim M$, ensuring that the dark matter correction remains subdominant relative to the black hole mass at astrophysically relevant scales, and observational constraints from stellar kinematics and lensing observations broadly support this range~\cite{Nampalliwar:2021tyz}. The event horizon(s) are determined by the zeros of \(\Delta(r)\), i.e. by the solutions of \(\Delta(r)=0\), namely
\begin{equation}
r^2+a^2-2m(r)\,r=0,
\end{equation}
which is generally transcendental because $m(r)$ contains $\ln r$.

\section{Null geodesics in the quantum improved Kerr black hole with PFDM} \label{S3}
The nonvanishing metric components for the above metric are expressed as:
\begin{align}
g_{tt} &= -\left(1-\frac{2m(r)r}{\Sigma}\right), \\
g_{t\varphi} &= -\frac{2a m(r) r\sin^2\theta}{\Sigma}, \\
g_{rr} &= \frac{\Sigma}{\Delta}, \qquad g_{\theta\theta}=\Sigma, \\
g_{\varphi\varphi} &= \sin^2\theta\left(r^2+a^2+\frac{2a^2 m(r) r\sin^2\theta}{\Sigma}\right).
\end{align}
The Lagrangian for photon motion is
\begin{equation}
\mathcal{L}=\frac12 g_{\mu\nu}\dot{x}^\mu \dot{x}^\nu ,
\end{equation}
where the overdot denotes differentiation with respect to an affine parameter.
The conjugate momenta are defined as
\begin{equation}
p_\mu = g_{\mu\nu}\dot{x}^\nu .
\end{equation}
Since the spacetime is stationary and axisymmetric, the conserved quantities are
\begin{equation}
p_t=-E, \qquad p_\varphi=L,
\end{equation}
where $E$ and $L$ represent the photon energy and angular momentum, respectively.
Solving for $\dot{t}$ and $\dot{\varphi}$, one obtains
\begin{align}
\dot{t} &= \frac{g_{\varphi\varphi}E+g_{t\varphi}L}{g_{t\varphi}^2-g_{tt}g_{\varphi\varphi}}, \\
\dot{\varphi} &= \frac{-g_{t\varphi}E-g_{tt}L}{g_{t\varphi}^2-g_{tt}g_{\varphi\varphi}}.
\end{align}
Using the identity
\begin{equation}
g_{t\varphi}^2-g_{tt}g_{\varphi\varphi}=\Delta\sin^2\theta,
\end{equation}
the above relations simplify to
\begin{align}
\dot{t} &= \frac{g_{\varphi\varphi}E+g_{t\varphi}L}{\Delta\sin^2\theta},
\label{eq:tdot} \\
\dot{\varphi} &= \frac{-g_{t\varphi}E-g_{tt}L}{\Delta\sin^2\theta}.
\label{eq:phidot}
\end{align}
Restricting to equatorial motion ($\theta=\pi/2$), the null condition
\begin{equation*}
g_{\mu\nu}\dot{x}^\mu\dot{x}^\nu = 0
\end{equation*}
leads to the radial equation of motion
\begin{equation}
\dot{r}^2 = \frac{1}{r^4}\left[(E(r^2+a^2)-aL)^2-\Delta(L-aE)^2\right].
\label{rdot_QI}
\end{equation}

\section{Null circular orbits and Lyapunov exponent} \label{S4}
\noindent Defining the radial potential as
\begin{equation}
R(r)=(E(r^2+a^2)-aL)^2-\Delta(L-aE)^2,
\end{equation}
the conditions for null circular orbits are
\begin{equation}
R(r_c)=0, \qquad \frac{dR}{dr}\Big|_{r=r_c}=0.
\end{equation}
Solving these equations, the critical impact parameter $b=L/E$ is obtained as
\begin{equation}
b_{\pm}(r_c)=\frac{a(2m(r_c)r_c)\pm r_c\sqrt{\Delta_c(2m(r_c)r_c)}}{2m(r_c)r_c-r_c^2},
\end{equation}
where $\Delta_c=\Delta(r_c)$ and the signs $\pm$ correspond to prograde and retrograde photon orbits.
The photon sphere radius satisfies
\begin{equation}
r_c^2-3m(r_c)r_c \pm 2a\sqrt{m(r_c)r_c}=0,
\label{photon_QI}
\end{equation}
which must be solved numerically due to the logarithmic PFDM contribution in $m(r)$.
The effective potential is defined as
\begin{equation}
V_{\rm eff}(r)=-\frac{1}{r^4}\left[(E(r^2+a^2)-aL)^2-\Delta(L-aE)^2\right].
\end{equation}
Expanding the radial potential around $r=r_c$,
\begin{equation}
R(r)\simeq \frac12 R''(r_c)(r-r_c)^2,
\end{equation}
the Lyapunov exponent governing the instability of null circular orbits is given by
\begin{equation}
\lambda_{\rm null}=
\sqrt{\frac{R''(r_c)}{2\,\dot{t}_c^{\,2}\,r_c^{4}}}.
\end{equation}
Using the explicit form of $\dot{t}_c$, the Lyapunov exponent reduces to
\begin{equation}
\lambda_{\rm null}=
\sqrt{
\frac{2\Delta_c-r_c^2(\Delta_c)''}
{2r_c^2(r_c^2+a^2-a b_c)^2}
}.
\label{lambda_QI}
\end{equation}
\begin{figure*}
	\centerline{
		\includegraphics[width=85mm,height=85mm]{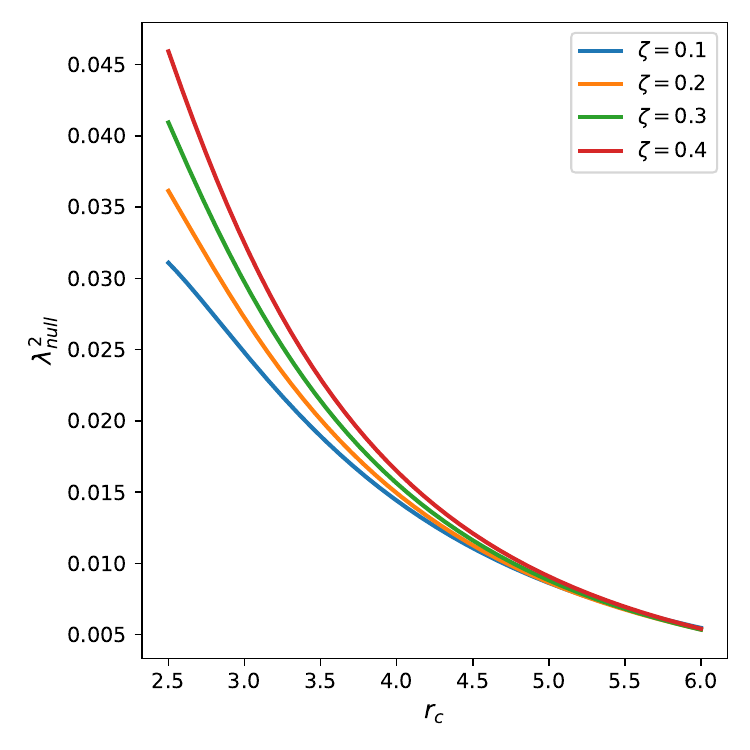}
        \includegraphics[width=85mm,height=85mm]{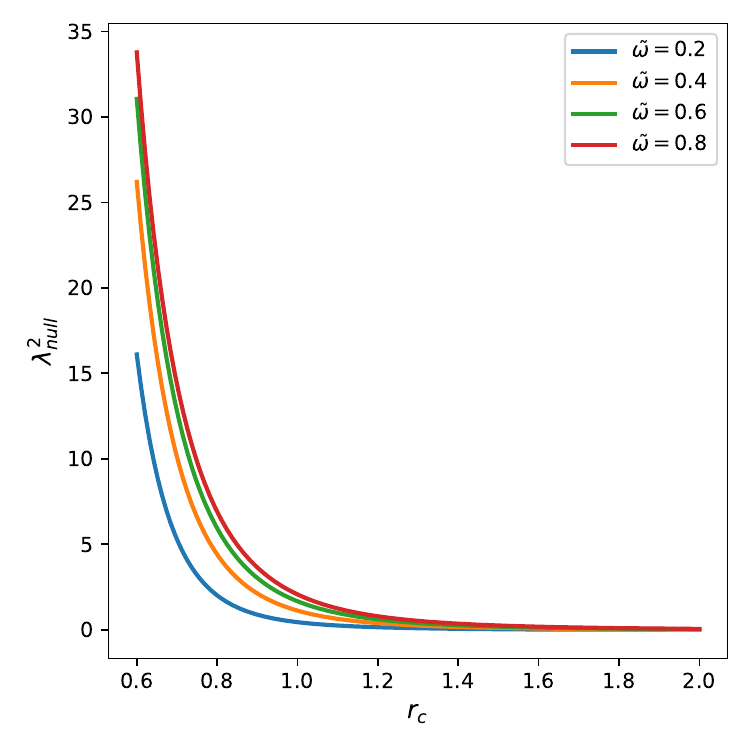}
        }
	\caption{Variation of Lyapunov exponent as a function of $r_{c}$ for different values of $\zeta$ and $\tilde{\omega}$. Here, we fix $M=1$, $a=0.4$, $\zeta=0.2$ and $\tilde{\omega}=0.4$, where applicable.}
\label{fig01}
\end{figure*}
A positive Lyapunov exponent indicates that the null circular orbits are dynamically unstable. Larger values of the null Lyapunov exponent correspond to stronger sensitivity to perturbations and enhanced chaotic behavior of photon trajectories.
In Fig.~\ref{fig01}, we plot the Variation of Lyapunov exponent as a function of $r_{c}$ for different values of $\zeta$ and $\tilde{\omega}$. The graphical representation clearly indicates that at fixed value of rotation, both parameter $\zeta$ and $\tilde{\omega}$ reduce the instability of black hole spacetime. However, as the critical radius of null orbits increases, the distinct values of both parameters coincide and saturated after a particular of $r_{c}$. Physically, this behavior suggests that the quantum improved parameter and PFDM framework significantly affect the instability of photon orbits. In the upcoming sections, we extend our analysis to measure the impact of parameters associated with black hole using advance stability indicators.  
\section{Poincaré Section} \label{S5}
Poincaré maps are a valuable tool in the study of nonlinear dynamical systems, providing a means to differentiate between order and chaos in phase-space~\cite{Cornish:1996de}. This method has been widely used to investigate the onset of chaotic behavior in the movement of test particles and photons within various black hole spacetimes~\cite{Guo:2022kio,Kumara:2024obd,Gallo:2024wju,Singh:2026vfd,Singh:2026odr}. However, the complete physical origins of chaos and its universal characteristics in intense gravitational fields are not yet fully comprehended. In this study, we utilize Poincaré sections to systematically examine the dynamic structure of geodesic motion around a quantum-improved rotating black hole surrounded by PFDM. Our main focus is on understanding how the quantum-improved parameter and PFDM contribution impact the stability of photon orbits and induce transitions between regular and chaotic behavior in phase space.\\
\begin{figure*}
	\centerline{
		\includegraphics[width=170mm,height=120mm]{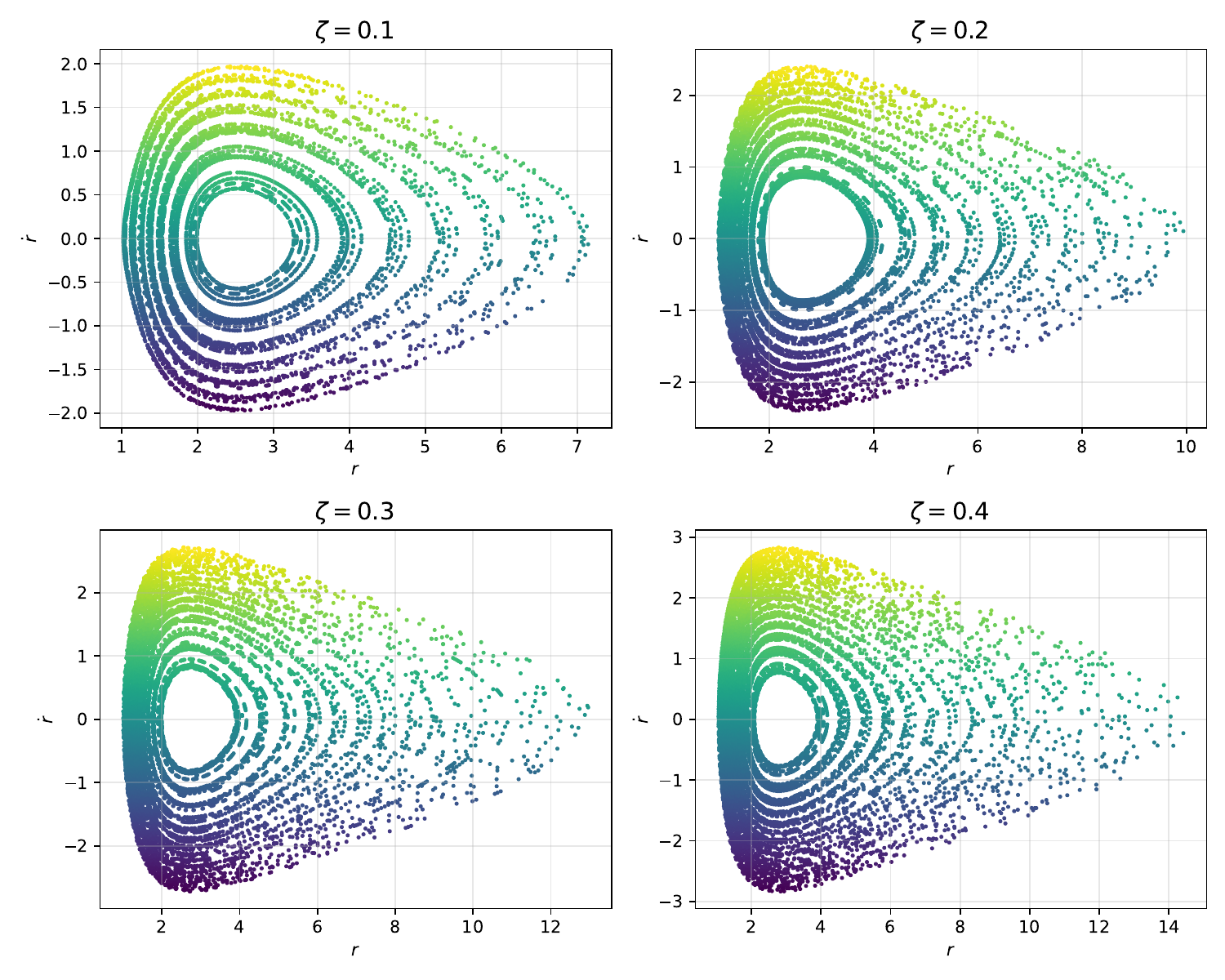}}
	\caption{Poincar\'e sections of null geodesic motion around the quantum-improved Kerr black hole surrounded by PFDM, constructed at $\theta=\pi/2$ with $p_{\theta}>0$, using multiple initial conditions. The PFDM parameter $\zeta$ varies across the panels, while the remaining parameters are fixed at $\tilde{\omega}=0.2$, $a=0.2$, $E=0.94$, and $L=4$.}
\label{fig1}
\end{figure*}
Fig.~\ref{fig1} presents the Poincar\'{e} sections of null geodesic motion in the $(r, \dot{r})$ phase plane for increasing values of the PFDM parameter $\zeta$. For the smallest value considered, the phase portrait is dominated by well-defined nested invariant tori surrounding a prominent central stability island, indicative of regular quasi-periodic photon motion. With increasing $\zeta$, the phase-space boundary expands progressively in radial extent, the central island becomes increasingly distorted, and the chaotic layer surrounding the regular region broadens systematically. These features confirm that the PFDM parameter enlarges the accessible phase-space volume and enhances orbital complexity through its growing contribution to the effective gravitational potential experienced by null geodesics.\\
\begin{figure*}
	\centerline{
		\includegraphics[width=170mm,height=120mm]{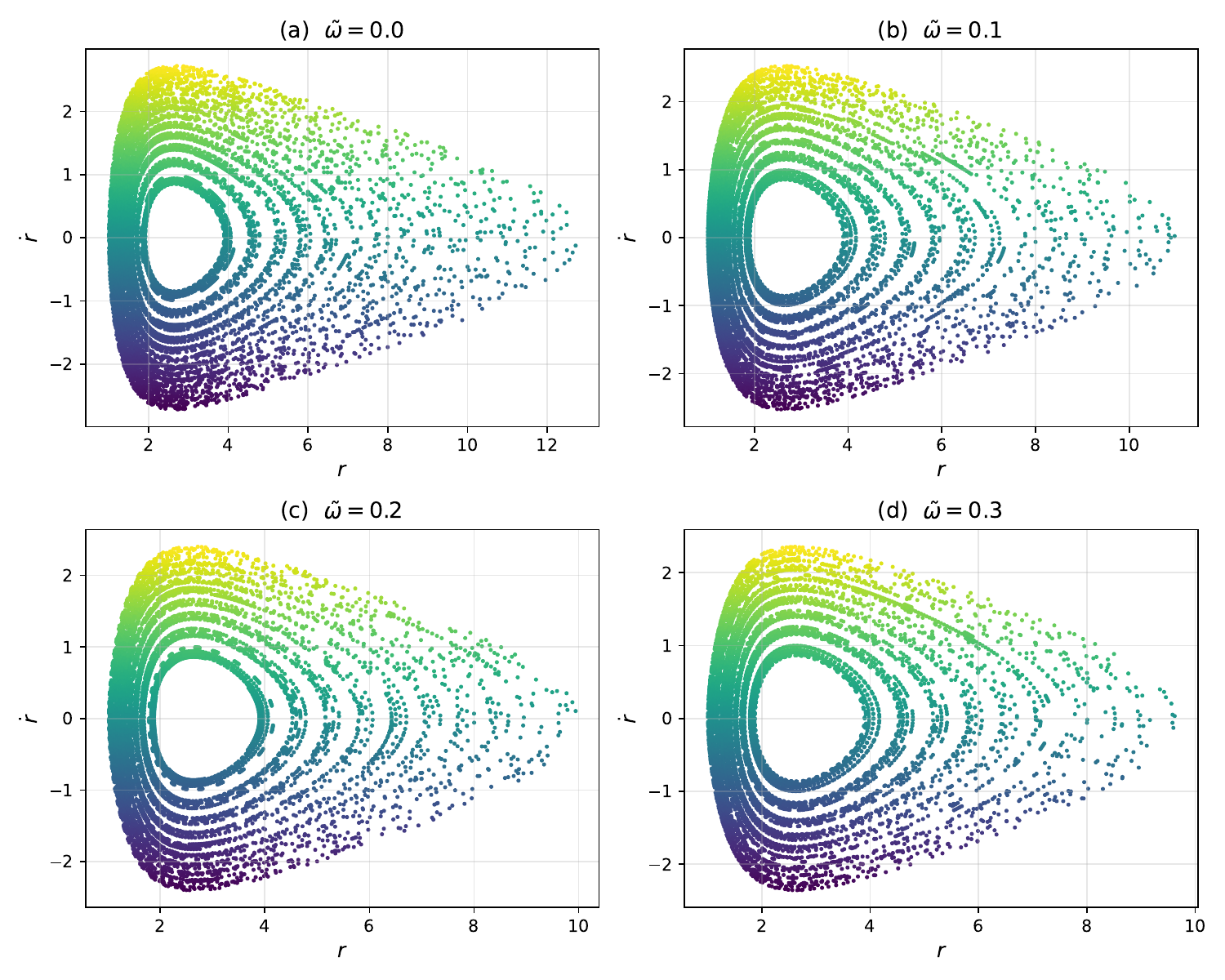}}
	\caption{Poincar\'e sections of null geodesic motion around the quantum-improved Kerr black hole surrounded by PFDM, constructed at $\theta=\pi/2$ with $p_{\theta}>0$, using multiple initial conditions. The quantum-improved parameter $\tilde{\omega}$ varies across the panels, while the remaining parameters are fixed at $\zeta=0.2$, $a=0.2$, $E=0.94$, and $L=4$.}
\label{fig2}
\end{figure*}
Fig.~\ref{fig2} displays the Poincar\'{e} sections for varying values of the quantum-improvement parameter $\tilde{\omega}$. In the absence of quantum corrections, the phase portrait exhibits densely packed nested tori characteristic of near-integrable geodesic dynamics. With increasing $\tilde{\omega}$, the radial extent of the section contracts noticeably, the central stable island becomes more compact, and the surrounding chaotic layer exhibits stronger phase-space mixing. This behavior reflects the role of the quantum-improvement parameter in modifying the near-horizon geometry, effectively strengthening gravitational confinement of null geodesics and 
reducing the radial range of stable photon orbits while simultaneously enhancing dynamical instability in the outer regions of the phase space.\\
\begin{figure*}
	\centerline{
		\includegraphics[width=170mm,height=120mm]{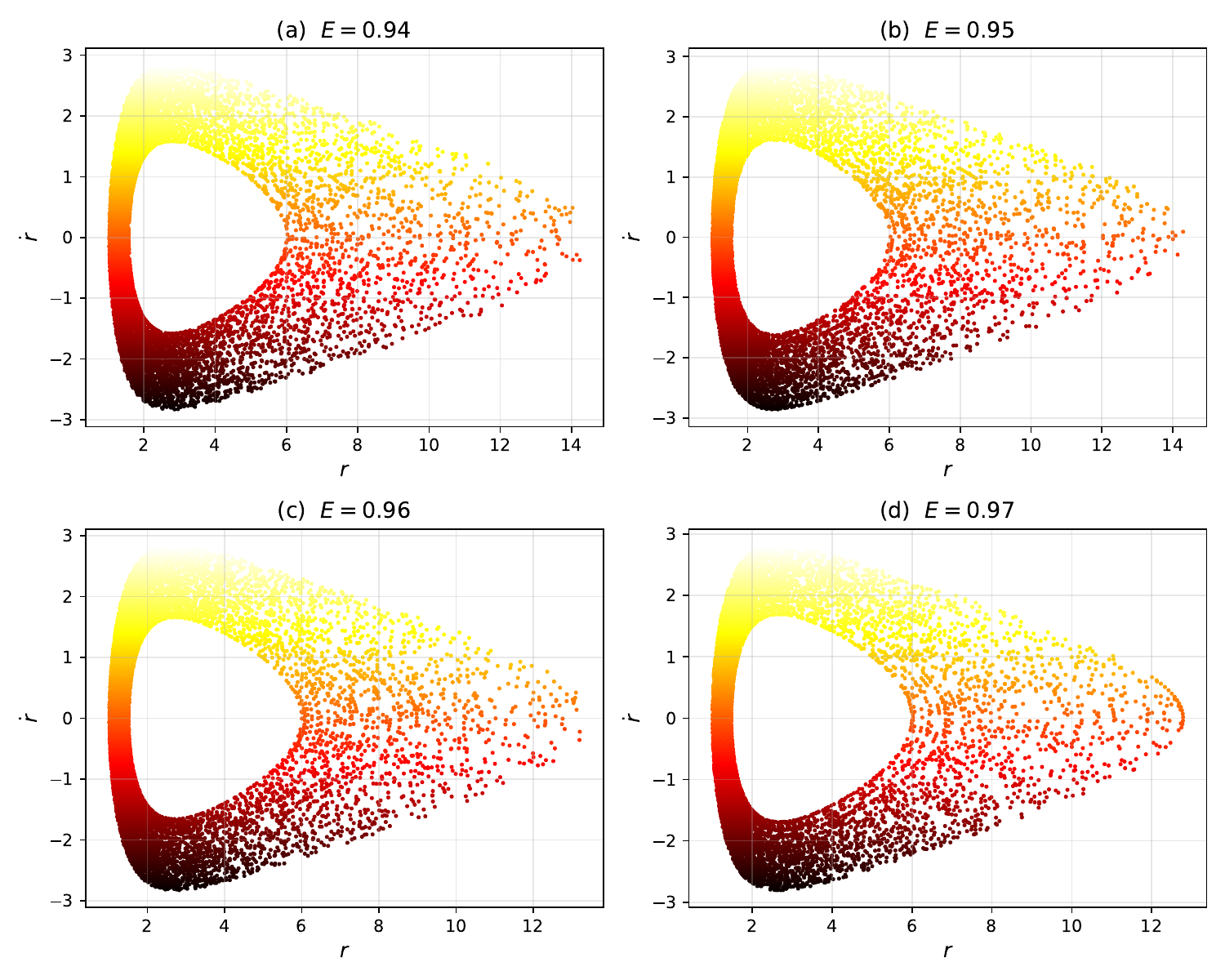}}
	\caption{Poincar\'e sections of null geodesic motion around the quantum-improved Kerr black hole surrounded by PFDM, constructed at $\theta=\pi/2$ with $p_{\theta}>0$, using multiple initial conditions. The energy parameter $E$ varies across the panels, while the remaining parameters are fixed at $\zeta=0.4$, $\tilde{\omega}=0.3$, $a=0.2$, and $L=4$.}
\label{fig3}
\end{figure*}
Fig.~\ref{fig3} shows the Poincar\'{e} sections for increasing values of the conserved photon energy $E$. The accessible phase-space region expands systematically in both the radial and momentum directions with increasing $E$, consistent with the greater kinetic energy available to the photon. The central stable island persists across all values examined but occupies a progressively smaller fraction of the total phase-space area, indicating that chaotic trajectories dominate an increasingly larger portion of the accessible region at higher energies. The outward shift of the phase-space boundary directly reflects the dependence of the outer radial turning point on the photon energy, establishing $E$ as the primary parameter governing 
the global extent of null geodesic motion in this spacetime. \\
Poincar\'e sections in the $(r,\dot{r})$ phase space obtained by numerically integrating the equations of motion given in Eqs.~\ref{eq:tdot}-\ref{rdot_QI} using the adaptive Runge--Kutta (RK45) method. The integrations are performed with relative and absolute tolerances of $10^{-8}$ and $10^{-10}$, respectively, over a sufficiently long integration interval to ensure that the phase-space structures are fully resolved. The Poincar\'e sections are constructed by recording successive crossings of the surface of section defined by $\phi=\pi/2$.
The Poincar\'{e} section analysis reveals that the PFDM parameter $\zeta$, the quantum-improvement parameter $\tilde{\omega}$, and the photon energy $E$ each exert a distinct and systematic influence on the phase-space structure of null geodesic motion in the quantum-improved Kerr spacetime. Increasing $\zeta$ and $E$ progressively enlarges the accessible phase-space region and enhances chaotic mixing, while increasing $\tilde{\omega}$ contracts the radial extent of photon orbits and strengthens gravitational confinement. Collectively, these results 
establish that the interplay between quantum gravitational corrections and dark matter contributions fundamentally governs the transition between regular and chaotic photon dynamics in this spacetime.
\section{Lyapunov Indicators} \label{S6}
The Largest Lyapunov Indicators (LLE) are important tools to study the stability of a dynamical system, as they reveal the sensitivity of photon path starting conditions and how quickly nearby paths in phase space diverge. The presence of strong gravitational fields deviate the photons from null circular geodesics and shift the regular motion of photon towards chaotic motion. In this subsection, we briefly study the LLE and their role in stability of null geodesics. Let us consider $\tau$ represent the proper time and $d(\tau)$ represent the distance between two nearby trajectories of null geodesics. In this physical setup, The LLE is defined as~\cite{Wu:2006rx,Han:2008zzf,Chen:2016tmr},
\begin{equation}
\lambda = \lim_{\tau \to \infty} \frac{1}{\tau} \ln\!\left(\frac{d(\tau)}{d(0)}\right),
\end{equation}
where $d(0)$ represents the initial separation. This formulation confirms that $\lambda$ is independent of coordinates and provides a measure of orbital instability in curved spacetime.\\ 
\begin{figure*}
	\centerline{
		\includegraphics[width=170mm,height=100mm]{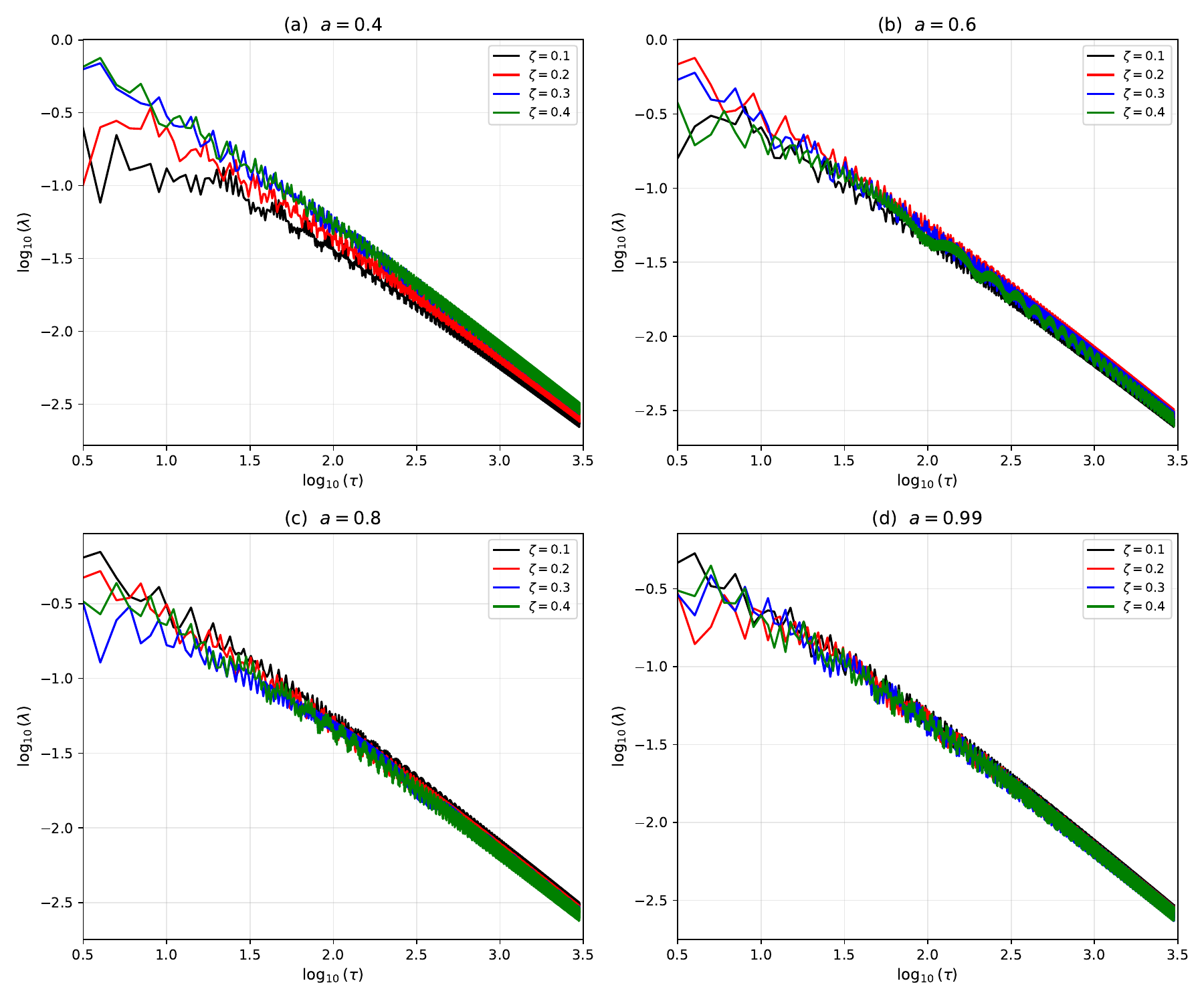}}
	\caption{Logarithmic Lyapunov exponent $\log_{10}\lambda$ versus $\log_{10}\tau$ for Quantum improved Kerr black hole surrounded by PFDM, shown for different values of the parameter $\zeta$. Panels (a)-(d) correspond to increasing values of the rotation parameter $a$. The remaining parameters are fixed at $\tilde{\omega}=0.2$, $E=0.95$, $L=3$ and $r_0 = 1.2$.}
\label{fig4}
\end{figure*}

\begin{figure*}
	\centerline{
		\includegraphics[width=170mm,height=100mm]{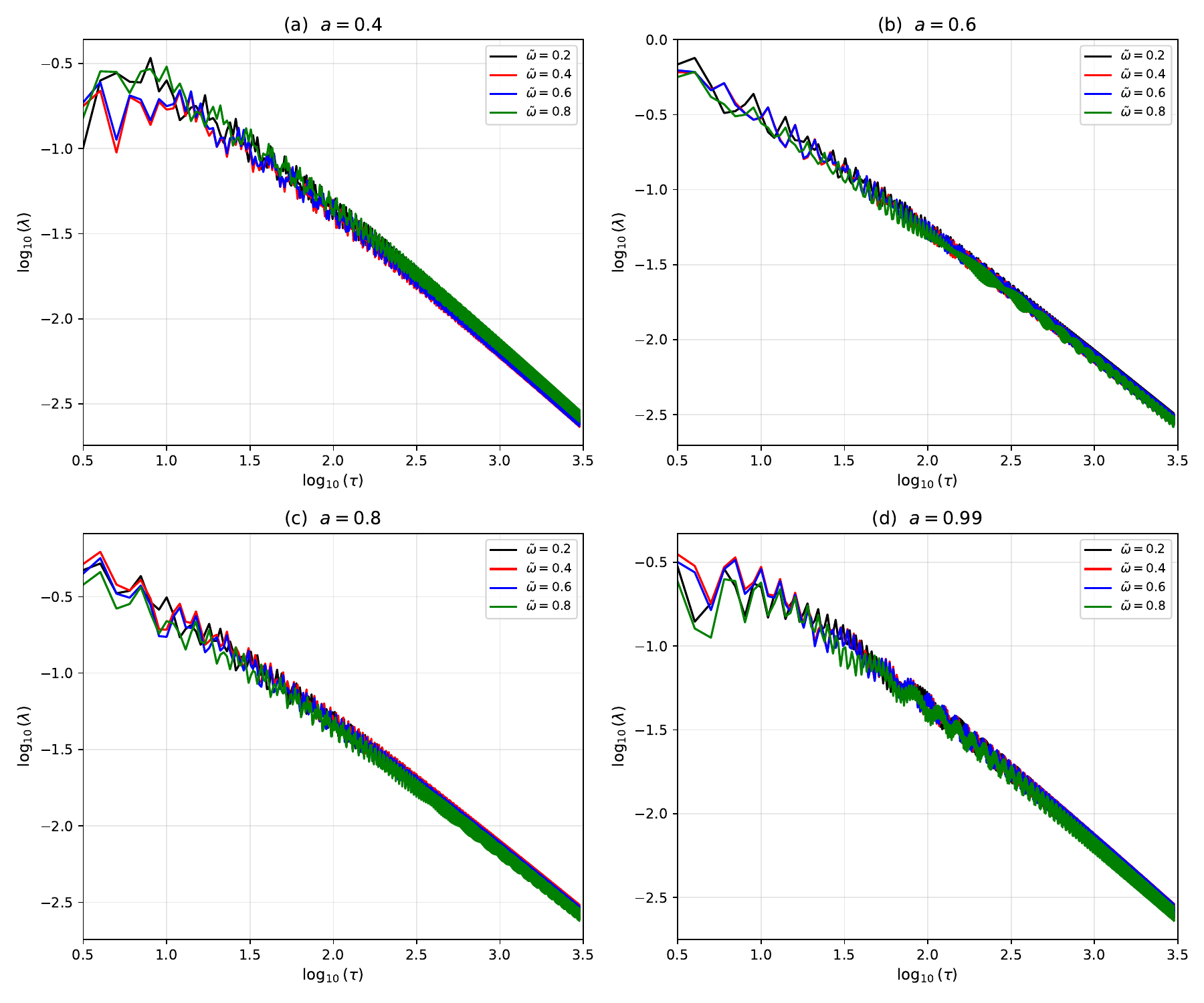}}
	\caption{Logarithmic Lyapunov exponent $\log_{10}\lambda$ versus $\log_{10}\tau$ for Quantum improved Kerr black hole surrounded by PFDM, shown for different values of the parameter $\tilde{\omega}$. Panels (a)-(d) correspond to increasing values of the rotation parameter $a$. The remaining parameters are fixed at $\zeta=0.2$, $E=0.95$, $L=3$ and $r_0 = 1.2$.}
\label{fig5}
\end{figure*}
In Figs.~\ref{fig4} and~\ref{fig5}, we plot the logarithmic Lyapunov exponent $\log_{10}\lambda$ versus $\log_{10}\tau$, which is commonly known as LLE, for a quantum-improved Kerr black hole surrounded by PFDM for different choices of parameters. In particular, the LLE for different values of the PDFM parameter with increased rotation is depicted in Fig.~\ref{fig4}. From the graphical analysis, we have observed that, at initial $\tau$, the separation among different values of PFDM is larger. It clearly indicates the PDFM parameter significantly affects the stability of photon motion in this phase space. In addition, as the $\tau$ increases, the $\log_{10}\lambda$ gradually moves towards the monotic decay, indicating the convergence in the asymptotic dynamical regime. Furthermore, the similar analysis for distinct values of quantum-improved parameters is represented in Fig.~\ref{fig5}. It has been observed that at initial $\tau$, the curves corresponding to different $\tilde{\omega}$ exhibit pronounced transient oscillations, suggesting short-time sensitivity to initial conditions and the influence of the local phase space structure. Notably, as the rotation increases, the sensitivity to the initial condition corresponding to the different $\tilde{\omega}$ reduces; therefore, transient oscillations are less dominant in this regime. Moreover, for all cases, as $\tau$ increases, the similar pattern of gradually monotonic decay in the asymptotic dynamics regime is observed. As compared to the PFDM parameter, the quantum-improved parameter is less dominant for later times. Thus, the analysis clearly shows that for a quantum-improved Kerr black hole in PFDM, the stability of photon motion has a significant dependence on the $\zeta$ and $\tilde{\omega}$, as they demonstrate how the regular motion is affected at the initial time during the separation of invariant curves. \\
\begin{figure*}
	\centerline{
		\includegraphics[width=170mm,height=100mm]{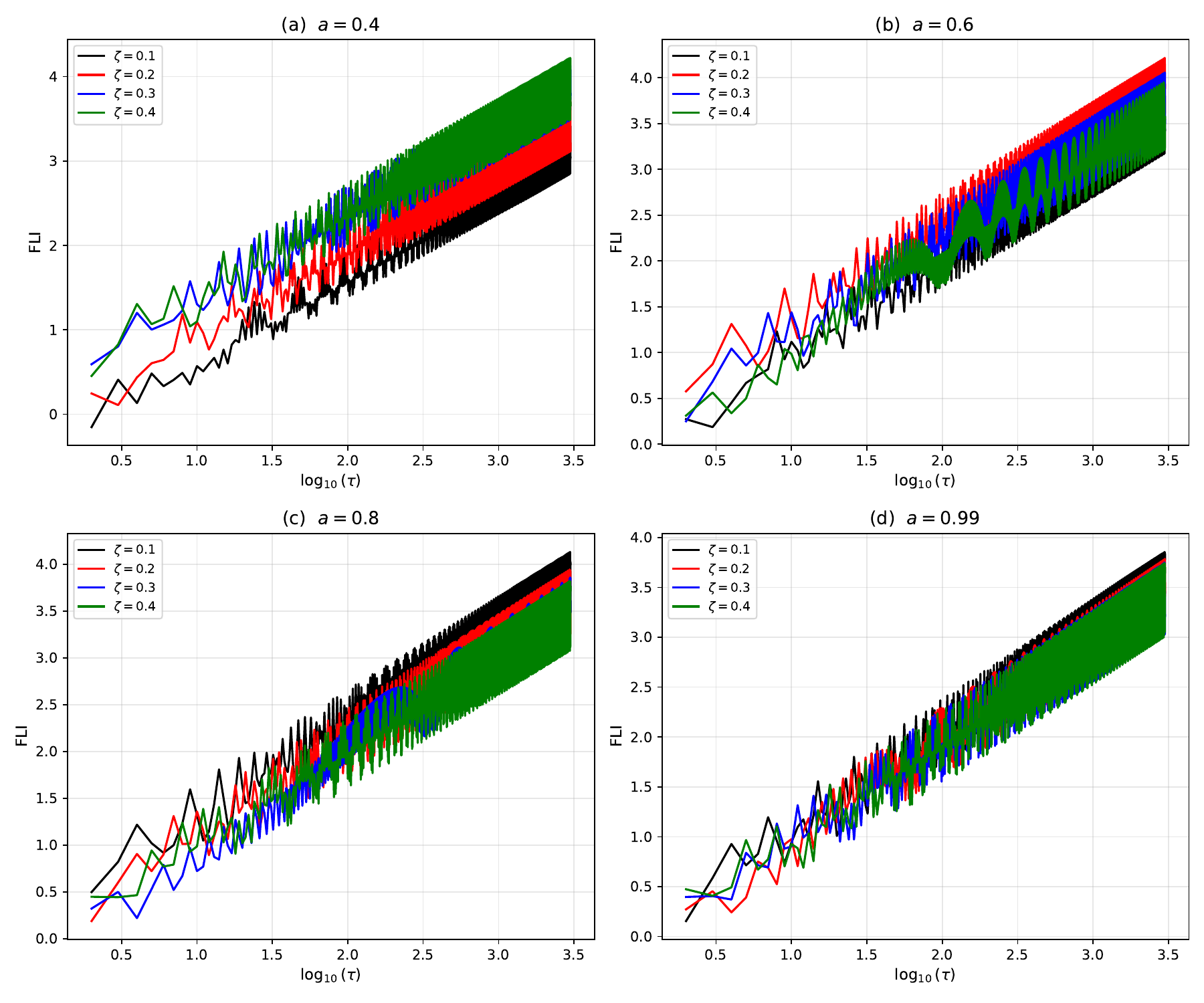}}
	\caption{The FLI versus proper time for Quantum improved Kerr black hole surrounded by PFDM, shown for different values of the parameter $\zeta$. From panel (a) to panel (d), the parameter sets $a$ are chosen to be increasing. The remaining parameters are fixed at $\tilde{\omega}=0.4$, $E=0.95$, $L=3$ and $r_0 = 1.2$. }
\label{fig6}
\end{figure*}
The Fast Lyapunov Indicators (FLI) are another important stability indicator as they provide robust techniques to examine the orbital stability characterization in phase space of Poincaré section. The FLI is very useful as it distinguishes the regular and chaotic orbits of much less duration if compared with the classical Lyapunov exponent. FLI tracks the rate of divergence of nearby photon orbits. 
To proceed further, let us consider $d(\tau)$ denote the norm of the deviation vector between two initially nearby geodesics at proper time $\tau$, with $d(0)$ representing their initial separation. The FLI is defined as~\cite{Wu:2006rx,Han:2008zzf,Chen:2016tmr},
\begin{equation}
\mathrm{FLI}(\tau) = \log_{10}\!\left(\frac{d(\tau)}{d(0)}\right).
\end{equation}
The variational equation for $d(\tau)$ can be derived from the geodesics of quantum-proved Kerr black hole in PFDM, ensuring that the FLI can be utilized to analyze the local instability features of photon trajectories caused by spacetime curvature. In the case of regular or quasiperiodic motion, the $\mathrm{FLI}(\tau)$ increases almost linearly with time, while there is a gradual increase when proper time is used as a reference. Conversely, chaotic trajectories exhibit exponential divergence, displaying strong sensitivity to initial conditions, leading to a significant increase in the FLI. The distinct behavior between regular, quasiperiodic and chaotic trajectories allows the time evolution of the FLI to serve as a precise measurement method, showcasing extreme sensitivity to minor perturbations caused by the strong curvature of spacetime, particularly affecting photon orbits. Consequently, the FLI serves as a tool to probe the transition from regular to chaotic motion of photons, as well as to identify the chaotic and stable regions within the phase space of rotating black holes in PFDM background.\\
\begin{figure*}
	\centerline{
		\includegraphics[width=170mm,height=100mm]{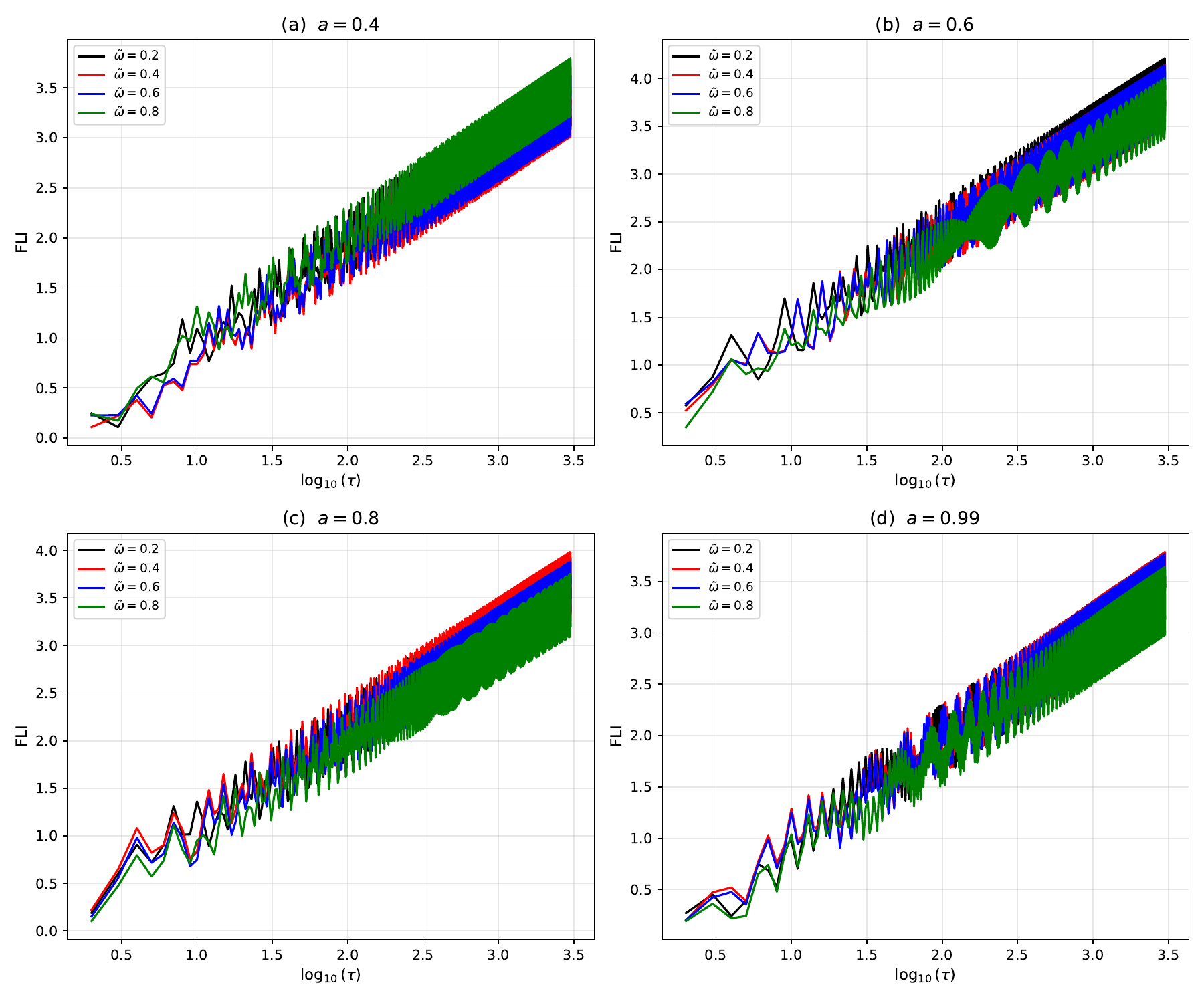}}
	\caption{The FLI versus proper time for Quantum improved Kerr black hole surrounded by PFDM, shown for different values of the parameter $\tilde{\omega}$. From panel (a) to panel (d), the parameter sets $a$ are chosen to be increasing. The remaining parameters are fixed at $\zeta=0.2$, $E=0.95$, $L=3$ and $r_0 = 1.2$. }
\label{fig7}
\end{figure*}
In Figs.~\ref{fig6} and~\ref{fig7}, we plot the FLI versus $\log_{10}\tau$ for a quantum-improved Kerr black hole surrounded by PFDM for different choices of parameters. Fig.~\ref{fig6} represents the FLI for different values of the PDFM parameter $\zeta$, with rotation increases from panel $(a)$ to $(d)$. The FLI analysis clearly shows that a systemic increase with the variation of $\tau$. However, at the initial time, the clear separation between different FLI curves corresponding to distinct $\zeta$ values was observed. Physically, this suggests that at the initial time, the stability of photon motion is very sensitive. Apart from the initial condition, the divergence becomes negligible, and the FLI curve becomes more compact, indicating the transformation from regular to chaotic motion. In particular, at lower values of rotation, the maximum FLI is observed for larger values of the PFDM parameter, and with increasing rotation, this systematic behavior changes due to the frame-dragging. Furthermore, in Fig.~\ref{fig7} we plot the FLI versus $\log_{10}\tau$ for different values of the quantum-improved parameter $\tilde{\omega}$. It has been observed that for all cases the FLI follows an increasing trend; this suggests the separation between photon trajectories exhibits a progressive divergence. At the initial time the separation between FLI curves indicates the sensitivity of photon motion for quantum-improved parameters in phase space. In particular, for all cases of rotation, the FLI curves reached such density for higher values of $\tilde{\omega}$. This clearly indicates the quantum-improved parameter significantly affects the motion of photons as the transition occurs from a regular to a chaotic regime. In summary, the analysis of FLI qualitatively demonstrates that the effect of PDFM and quantum-improved parameters plays a significant role on the stability of photon motion in this black hole spacetime. 
\section{Kolmogorov--Sinai Entropy} \label{S7}
The Kolmogorov-Sinai (KS) entropy is a rigorous quantitative measure of chaoticity in a dynamical system. It captures the rate at which information about a trajectory is produced or lost over time in phase space. In Hamiltonian systems, the KS entropy is closely connected to the Lyapunov spectrum via Pesin's theorem. Pesin's theorem states that the KS entropy equals the sum of all positive Lyapunov exponents~\cite{pesin1977characteristic},
\begin{equation}
h_{\mathrm{KS}} = \sum_{\lambda_i > 0} \lambda_i.
\end{equation}
In this study, the KS entropy is calculated based on the largest finite-time Lyapunov exponent, which is determined by analyzing the divergence of initially nearby null geodesics~\cite{Pradhan:2015aaa,Mondal:2021exj}. Positive values of $h_{ \mathrm{KS}}$ indicate chaotic motion, demonstrating exponential sensitivity to initial conditions, while $h_{\mathrm{KS}} \approx 0$ corresponds to regular, quasi-periodic trajectories confined to invariant tori. Physically, $h_{\mathrm{KS}}$ quantifies the exponential rate at which initially neighboring photon trajectories diverge in the curved spacetime of a rotating black hole in Weyl conformal gravity. As a result, it offers a direct measure of the inherent unpredictability in photon dynamics under strong gravitational fields. Higher KS entropy values signify a faster loss of predictability and stronger mixing in phase space, whereas lower values indicate more ordered motion. By computing $h_{\mathrm{KS}}$ across a variety of initial conditions and black hole parameters, researchers can delineate regions of regularity and chaos, leading to a deeper understanding of the structure and stability of photon orbits within this conformal gravity background.\\
\begin{figure*}
	\centerline{
		\includegraphics[width=170mm,height=220mm]{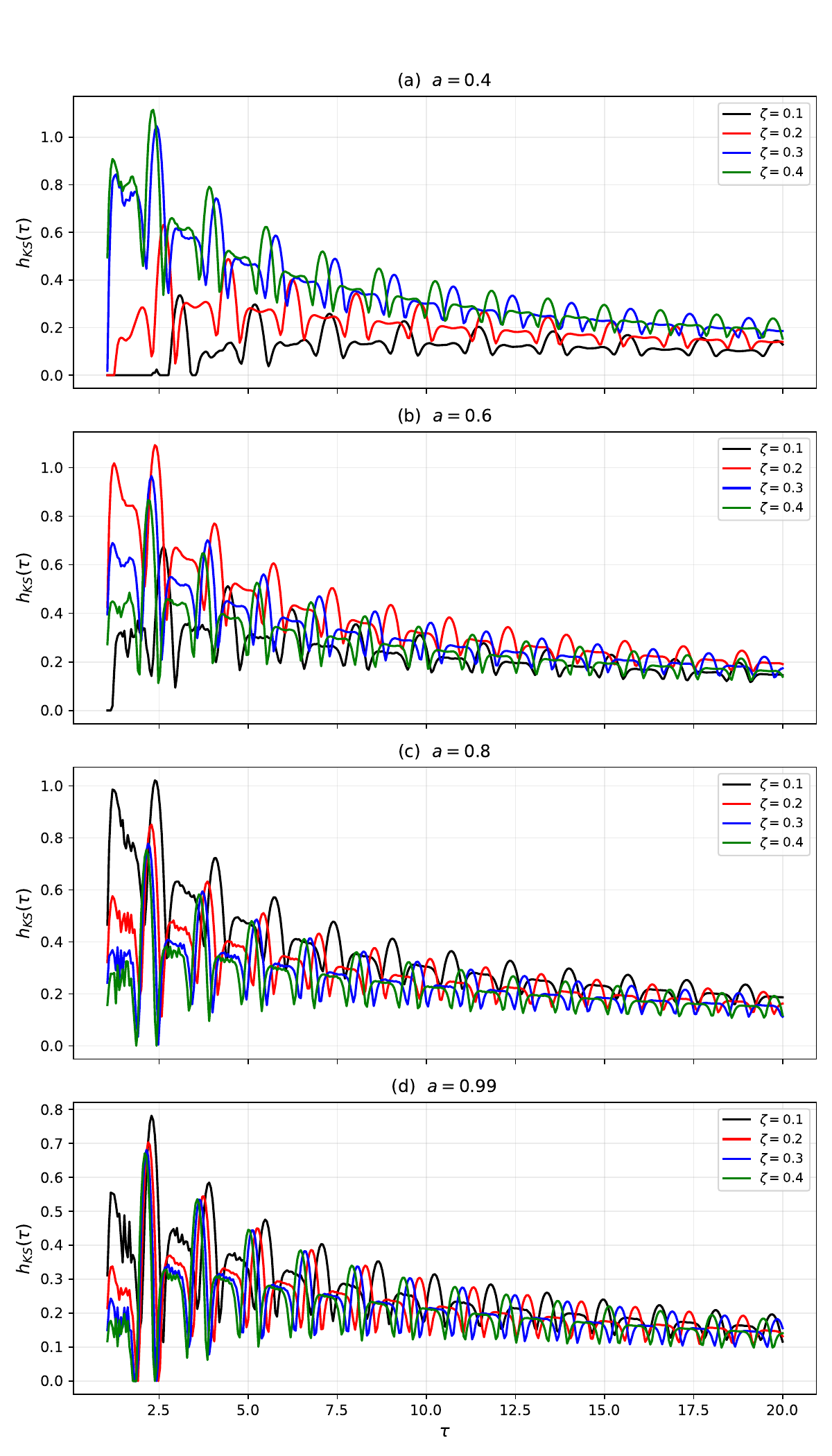}}
	\caption{Time evolution of the Kolmogorov--Sinai (KS) entropy $h_{KS}(\tau)$ for null geodesics of Quantum improved Kerr black hole surrounded by PFDM, shown for different values of the parameter $\zeta$. The panels from top to bottom show increasing values of the rotation parameter $a$. The remaining parameters are fixed at $\tilde{\omega}=0.4$, $E=0.95$, $L=3$ and $r_0 = 1.2$.
}
\label{fig8}
\end{figure*}
In Fig.~\ref{fig8} we plot the time evolution of the KS entropy $h_{KS}(\tau)$ for null geodesics of a quantum-improved Kerr black hole surrounded by PFDM for different values of the parameter $\zeta$. The panels from top to bottom show increasing values of the rotation parameter $a$. The remaining parameters are fixed at $\tilde{\omega}=0.4$, $E=0.95$, $L=3$, and $r_0 = 1.2$.
The analysis shows that at initial times the pronounced oscillatory behavior is exhibited, suggesting the high sensitivity of photon trajectories corresponding to initial conditions. Further, the amplitude of oscillatory motion gradually decreases as the time increases, indicating the transition towards weak chaotic and regular motion. In addition, one can see that, for a fixed rotation parameter, the magnitude as well as the damping rate of entropy oscillation increases with an increase in the PFDM parameter. It clearly demonstrates that the presence of dark matter backgrounds significantly affects the stability and chaotic properties of photon motion. Moreover, as we compare the role of different rotation parameters, it is observed that for maximum rotation, the initial time entropy profile seems more complex and irregular. This physically indicates that with the presence of the PFDM parameter, the higher rotation produces stronger fluctuation and slower decay in KS entropy. Overall, one can summarize that the presence of the PFDM parameter for different rotations at a fixed value of the quantum-improved parameter plays a crucial role in regulating the degree of chaos, hence the long-term stability of null geodesics in the black hole as well.  \\
\begin{figure*}
	\centerline{
		\includegraphics[width=170mm,height=220mm]{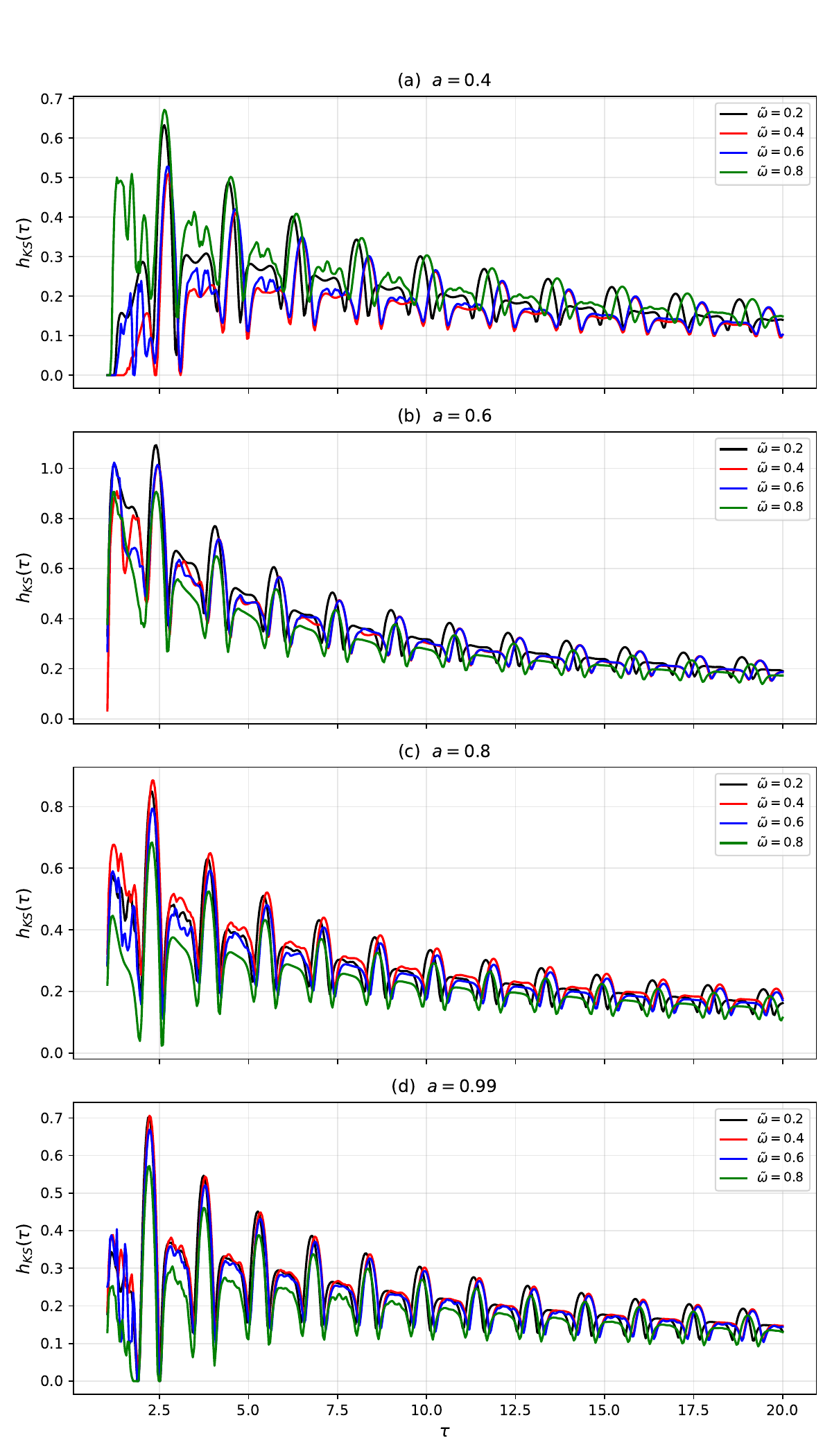}}
	\caption{Time evolution of the Kolmogorov--Sinai (KS) entropy $h_{KS}(\tau)$ for null geodesics of Quantum improved Kerr black hole surrounded by PFDM, shown for different values of the parameter $\tilde{\omega}$. The panels from top to bottom show increasing values of the rotation parameter $a$. The remaining parameters are fixed at $\zeta=0.2$, $E=0.95$, $L=3$ and $r_0 = 1.2$.}
\label{fig9}
\end{figure*}
The KS entropy analysis for different values of the quantum-improved parameter $\tilde{\omega}$ at different rotations is presented in Fig.~\ref{fig9}. For all cases, the KS entropy has a change in its behavior at the initial time, and then it slowly settles down over time. In addition, the amplitude of oscillatory motion gradually decreases as the time increases, indicating the transition towards weak chaotic and regular motion, similar to the previous analysis. For lower rotation, the KS entropy is observed to be maximum for the larger value of the quantum-improved parameter. However, as we increase rotation from $a=0.4$, the KS entropy produces a minimum value for the higher, larger quantum-improved parameter. This sudden change in the profile of KS entropy clearly depicts the crucial role of the combination of rotation and quantum-improved parameter in the chaotic dynamics of this black hole spacetime. If we combine the effects of rotation and the quantum-improved parameter $\tilde{\omega}$, we see some differences from what we would expect from the classical Kerr behavior, especially when the gravity is very strong. These results demonstrate that the quantum corrections and PFDM background in Kerr spacetime play a significant role for controlling the chaos near the black hole horizon. 
\section{Quantitative Detection of Chaos Using Weighted Birkhoff Averages} \label{S8}
Poincaré sections and Lyapunov exponents are qualitative diagnostics used to identify chaotic dynamics. In this analysis, the Weighted Birkhoff Average (WBA) is employed as a robust numerical technique to distinguish between regular and chaotic photon trajectories in the quantum-improved rotating black hole surrounded by PFDM. This method utilizes the convergence properties of time-averaged observables along individual orbits, enabling an efficient classification of regular versus chaotic motion~\cite{eckmann1985ergodic,cornfeld2012ergodic,das2016measuring,meiss2021birkhoff,sander2020birkhoff,duignan2023distinguishing,Kala:2026qej}.
\begin{figure*}
	\centerline{
		\includegraphics[width=170mm,height=100mm]{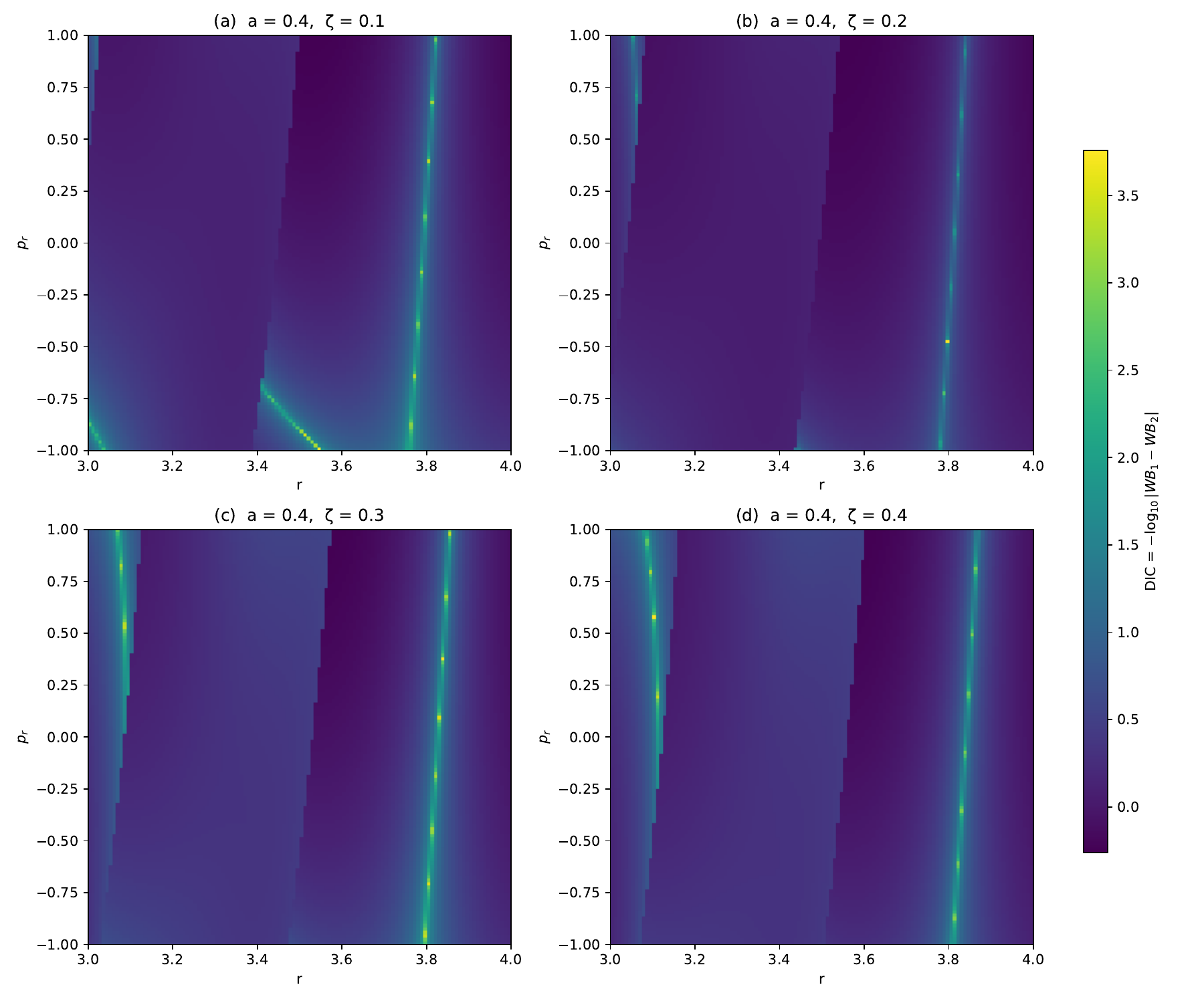}}
	\caption{DIC maps in the $(r,p_r)$ phase space for null geodesics of Quantum improved Kerr black hole surrounded by PFDM for different values of $\zeta$. The color bar represents the DIC value $-\log_{10}|WB_1-WB_2|$. The parameters are fixed at $a=0.4$, $\tilde{\omega} = 0.2$, $E=0.95$ and $L = 3$.}
\label{fig10}
\end{figure*}
\begin{figure*}
	\centerline{
		\includegraphics[width=170mm,height=100mm]{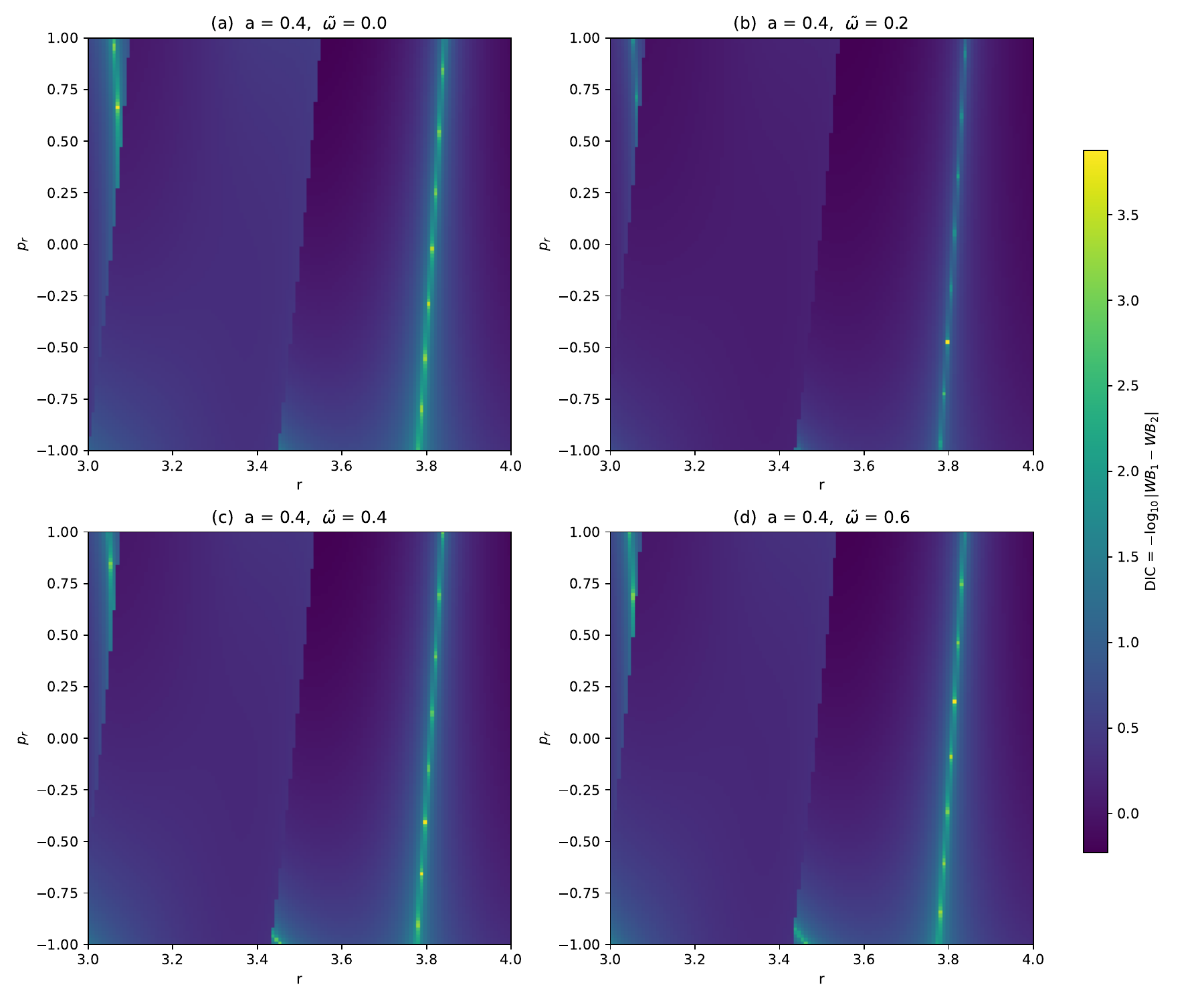}}
	\caption{DIC maps in the $(r,p_r)$ phase space for null geodesics of Quantum improved Kerr black hole surrounded by PFDM for different values of $\tilde{\omega}$. The color bar represents the DIC value $-\log_{10}|WB_1-WB_2|$. The parameters are fixed at $a=0.4$, $\zeta = 0.2$, $E=0.95$ and $L = 3$.}
\label{fig11}
\end{figure*}
For equatorial null geodesics ($\theta=\pi/2$) with $p_\theta>0$, the reduced phase space is characterized by the canonical coordinates $(r,p_r)$. Each photon orbit intersects a prescribed Poincaré surface of section multiple times, generating a discrete sequence of points that defines an associated iterated map
\begin{equation}
\mathbf{v}_{n+1} = \mathcal{M}(\mathbf{v}_n),
\end{equation}
where $\mathcal{M}$ is the Poincaré return map associated with the continuous geodesic flow.
Along this discrete orbit, a smooth observable is defined as
\begin{equation}
f(r,p_r) = \sin\!\big[2\pi(r+p_r)\big],
\end{equation}
which has been proven effective in discriminating between different motion types in dynamical systems. The Weighted Birkhoff Average of $f$ over $N$ successive iterations is given by a summation involving normalized weights and the observable function, by~\cite{cornfeld2012ergodic,das2016measuring}
\begin{equation}
WB_N(f)(\mathbf{v}_0) = \sum_{n=0}^{N-1} w_{n,N} \, f\!\left(\mathcal{M}^n(\mathbf{v}_0)\right),
\end{equation}
with normalized weights
\begin{equation}
w_{n,N} = \frac{g(n/N)}{\sum_{k=0}^{N-1} g(k/N)}.
\end{equation}
The weighting function is constructed using a smooth bump profile~\cite{eckmann1985ergodic}:
\begin{equation}
g(z)=
\begin{cases}
\exp\!\left[-\dfrac{1}{z(1-z)}\right], & 0<z<1, \\
0, & \text{otherwise},
\end{cases}
\end{equation}
which mitigates boundary effects and accelerates convergence for regular trajectories, while chaotic orbits exhibit slower, irregular convergence. To quantify the convergence behavior, we compute two WBAs from successive segments of the same trajectory. Their difference defines a DIC~\cite{eckmann1985ergodic,cornfeld2012ergodic},
\begin{equation}
\mathrm{DIC} = -\log_{10}\!\left| WB_N^{(1)} - WB_N^{(2)} \right|.
\end{equation}
The large DIC values indicate rapid convergence and regular (quasi-periodic) trajectories, whereas small DIC values signal chaotic behavior. 
The DIC is computed on a dense grid of initial conditions in $(r,p_r)$ phase space to produce two-dimensional color maps of the global dynamics, revealing regular islands, chaotic layers, and their interfaces. This WBA-based DIC method provides a fast, numerically stable alternative to Lyapunov-based diagnostics, well-suited for high-resolution scans of black hole phase spaces. For the visualizations, a Python implementation inspired by Ref.~\cite{rolim2025pynamicalsys} is utilized. Null geodesics are evolved up to $T=200$ with $N=4000$ sampling points per trajectory, and the phase space is sampled on a uniform lattice with $N_r=120$ radial and $N_{p_r}=120$ momentum points. For each initial condition $(r,p_r)$, the orbit is integrated, the WBAs are computed, and the DIC is obtained from differences between successive averages, yielding global DIC maps that sharply highlight transitions between ordered and chaotic regimes as system parameters vary.\\

Fig.~\ref{fig10} presents the Dynamical Indicator of Chaos (DIC) maps in the $(r,p_r)$ phase space for null geodesics around the quantum-improved Kerr black hole surrounded by PFDM, for different values of the PFDM parameter $\zeta$, with fixed parameters $a=0.4$, $\tilde{\omega}=0.2$, $E=0.95$, and $L=3$. The color scale represents the DIC defined as $-\log_{10}|WB_1-WB_2|$, where darker (purple) regions correspond to regular and stable motion, while brighter (yellow--green) regions indicate strong sensitivity to initial conditions and the onset of chaotic dynamics. As $\zeta$ increases from panel (a) to (d) the structure in phase space changes a lot. It has been observed that, the bright chaotic bands become clearer as move to larger radial distances. This change shows that the PFDM parameter has an effect on how photons move. It makes some areas in phase space more unstable. The PFDM parameter changes the potential that governs photon motion.
This results in pronounced chaotic bands. The phase-space structure exhibits deformation. The PFDM parameter significantly modifies the potential. It enhances instability, in specific regions of phase space. The persistence of extended dark regions suggests that regular or quasi-periodic photon orbits still survive; however, their domain progressively shrinks with increasing $\zeta$. Hence, PFDM plays a crucial role in controlling the balance between stable and chaotic null trajectories.\\

Fig.~\ref{fig11} shows the DIC maps in the $(r,p_r)$ phase space for null geodesics of the quantum-improved Kerr black hole surrounded by PFDM for different values of the quantum improvement parameter $\tilde{\omega}$, while keeping $a=0.4$, $\zeta=0.2$, $E=0.95$, and $L=3$ fixed. With increasing $\tilde{\omega}$, the phase-space pattern changes appreciably and the chaotic regions become more structured and extended. This reflects the influence of quantum gravitational corrections on photon dynamics, leading to stronger sensitivity to initial conditions and increased instability of null orbits. If we look at this next to Fig.~\ref{fig10}, the effect of $\tilde{\omega}$ is not as obvious. It is still there, in a regular way. This shows that the quantum corrections are changing the details of phase space rather than causing sudden changes to chaos. The quantum corrections are really just tweaking the structure of phase space. The coexistence of regular and chaotic regions confirms the mixed dynamical nature of null geodesics in this quantum-corrected rotating spacetime with PFDM background.

\section{Conclusion and Discussion} \label{S9}
We have investigated the nonlinear photon dynamics in quantum improved rotating black hole surrounded by perfect fluid dark matter (PFDM) using various methods of analysis, including Poincaré sections, Lyapunov exponents, Kolmogorov-Sinai (KS) entropy and weighted Birkhoff averages (WBA). We first examined the behavior of Lyapunov exponent and obtained that the presence of quantum-improved parameter $(\tilde{\omega})$ and PFDM parameter $(\zeta)$, both reduce the instability of black hole spacetime. Further, we present a comprehensive analysis of the phase space trajectories using Poincar\'{e} maps. The Poincar\'{e} section analysis demonstrates that the PFDM parameter $\zeta$, the quantum-improvement parameter $\tilde{\omega}$, and the photon energy $E$ each exert a distinct and systematic influence on the phase-space structure of null geodesic motion in the quantum-improved Kerr spacetime. Increasing $\zeta$ and $E$ progressively enlarges the accessible phase-space region and enhances chaotic mixing, while increasing $\tilde{\omega}$ contracts the radial extent of photon orbits and strengthens gravitational confinement. In addition, these parameters significantly influence the structure of invariant tori in phase space by modifying their size and distribution, thereby affecting the stability of photon trajectories. Next, we use the Lyapunov indicators to further examine the qualitative behavior of the stability of photon motion. The logarithmic Lyapunov exponent (LLE) and Fast Lyapunov Indicator (FLI) show that photon motion stability in a quantum-improved Kerr black hole surrounded by PFDM depends greatly on the $\zeta$ and $\tilde{\omega}$. Larger values of $\zeta$ modify the early-time separation of trajectories, showing stronger sensitivity to initial conditions, while $\tilde{\omega}$ mainly affects transient dynamics and becomes less dominant later. Both indicators show a gradual monotonic decay in the asymptotic regime, signaling convergence toward regular motion. LLE and FLI distinguish regular from chaotic photon trajectories and show how these parameters control the shift between stability and chaos in phase space. Furthermore, the KS entropy analysis shows strong initial oscillations that gradually damp with time, indicating a transition from chaotic to more regular photon motion, with both quantum-improved parameter and the PFDM parameter significantly influencing the amplitude and decay rate of these oscillations. Their combined interplay with the rotation parameter governs the degree of chaos, showing that dark matter effects and quantum corrections critically control the long-term stability of null geodesics near the black hole. Finally, we use the WBA method which offers a robust quantitative criterion to distinguish regular and chaotic orbits and enables detailed mapping of the phase-space structure. The graphical representation is shown through the DIC map. The DIC maps show that increasing the quantum-improved parameter enhances and spreads chaotic regions in phase space, shrinking the domain of regular photon orbits and indicating stronger instability of null trajectories. Increasing the PFDM parameter changes the phase-space structure more smoothly, adding structured chaotic features without sudden shifts. This shows the subtle yet systematic impact of quantum corrections on photon dynamics. In summary, the results of the present analysis have direct implications for several astrophysical observables associated with the quantum-improved Kerr black hole in the PFDM environment.  Since the instability of photon orbits directly governs the observable optical properties of black holes, our results offer a mathematical framework to probe the imprints of quantum corrections and dark matter effects in future high-precision astronomical observations. In principle, these results could also bear on QPO-related observables, since the parameter-dependent modification of fundamental photon orbital frequencies implies shifts in associated oscillation timescales that may in principle be detectable in the variability of emission from black hole accretion environments, offering an additional observational window on the quantum gravitational and dark matter effects characterized in this work.

\section*{Acknowledgments}
This research was funded by the National Natural Science Foundation of China (NSFC) under Grant No. U2541210 and 12505238. The Work of S. Saghafi is financially supported by the INSF of  Iran under the grant number 40408889.  

\bibliographystyle{unsrt}
\bibliography{mainKBHPFDMQI}

\end{document}